\documentclass[%
 reprint,
showpacs,
 amsmath,
 amssymb,
 aps,
 pre,
floatfix,
]{revtex4-1}

\usepackage{graphicx}
\usepackage{dcolumn}
\usepackage{bm}

\begin{document}

%\preprint{APS/123-QED}
\title{Intermittent direction reversals of moving spatially-localized turbulence observed in two-dimensional Kolmogorov flow}

\author{Yoshiki HIRUTA}
\email[]{hiruta@kyoryu.scphys.kyoto-u.ac.jp}
\author{Sadayoshi TOH}
\affiliation{%
  Division of Physics and Astronomy, Graduate School of Science, Kyoto University, Japan}%

\date{\today}
\begin{abstract}
 We have found that in two-dimensional Kolmogorov flow
 a single spatially-localized turbulence (SLT) exists stably and
 travels with a constant speed on average  switching the moving
 direction randomly and intermittently for moderate values of
 control parameters: Reynolds number and the flow rate.
 We define the coarse-grained  position and velocity of an SLT and
 separate the motion of the SLT from its internal turbulent dynamics
 by introducing a co-moving frame.
 The switching process of an SLT represented by the coarse-grained
 velocity seems to be a random telegraph signal.
 Focusing on the asymmetry of the internal turbulence
 we introduce two coarse-grained variables characterizing the internal
 dynamics. These quantities follow the switching process reasonably.
 This suggests that the twin attracting invariant sets each of which
 corresponds to a one-way traveling SLT are embedded in the attractor
 of the moving SLT and the connection of the two sets is too
 complicated to be represented by a few degrees of freedom
 but the motion of an SLT is controlled  by the internal
 turbulent dynamics.
\end{abstract}

\pacs{}
\maketitle
\section{introduction}
Spatially-localized turbulent states (SLT) embedded in laminar flows such as puff and stripe,
are observed mainly in subcritical transient flows around nonlinear critical Reynolds number
both experimentally and numerically\cite{Avila2011,Shimizu2014,Willis2008,Khapko2016,Duguet2010,Ishida2016}.
These SLTs play a fundamental role in elucidation of generation, evolution
and sustenance of turbulence as well as transition to turbulence.

Considering not globally-occupied but spatially-localized states,
new aspects of turbulence are emerged such as motion of turbulent regions.
Since turbulence states  are localized, 
the position and  velocity of a turbulent state can be defined.
Furthermore, these facts  may stimulate researchers in more general
context such as dissipative
soliton and self-propelled particle: the former is a  moving solitary
state in a dissipative system  and the latter is a simple model of animate
lives such as microorganism, bird, fish and their collective motion.

In these contexts, a spatially-localized turbulent state can be regarded
as a moving element coupled with complex internal freedoms.
These moving turbulent regions also are connected with phenomena
interfering with our daily life.
For example, typhoons, which cause severe disasters, are
 fully developed complex turbulence and needless to say,
 prediction of their paths is not still easy.

 Collective behavior of SLTs plays also an essential role
in subcritical transitions. 
In such transient flows,  SLTs  create their copies and annihilate stochastically \cite{Avila2011}.
Recently,  experimental and numerical researches  have uncovered 
that subcritical transitions in shear flows  can be regarded
as the absorbing phase transition and its scaling exponents
accord with those of directed percolation\cite{Sano2015,Lemoult2016a}.

To describe the dynamics of complex turbulent states,
 dynamical systems approaches have been widely applied nowadays.
 In these approaches, simple invariant solutions of
  governing equations such as periodic solutions 
are adopted  as landmarks embedded in a phase space,
and a certain realization is identified as a single trajectory visiting
these unstable invariant solutions
\cite{Nagata1990,Avila2013,Willis2013,Kreilos2012a,Kreilos2013,Mellibovsky2012,Kawahara2012,Kawahara2001,Itano2001,Chandler2013,Lucas2014c}.
Numerical methods to find unstable solutions  based on Newton method have been developed to obtain a good guess for dealing with complex
flows even at relatively higher Reynolds
 number\cite{Teramura2014,Farazmand2015}. This approach
 has been extended to the results of laboratory experiments \cite{Suri2017}.
However, it is a hard task to research dynamical properties of
spatially-localized states because we must treat a wide range of
spatial modes from small ones representing turbulence 
to large ones isolating turbulence from laminar regions.

While dynamical and statistical properties of flows in relatively small
systems at low or moderate Reynolds numbers have been well understood,
those of turbulent flows in extended domains at higher Reynolds number
are not still clarified. 
This is partially because inhomogeneity induced by walls plays
a crucial role in developed wall-bounded  flows.
In fact,  many ingredients of turbulent flows including near-wall
dynamics and large scale structures in bulk
spontaneously coexist and interact with each other \cite{Toh2005}.
In addition, dynamical description of systems with translational
symmetries has been studied  for a long time\cite{Budanur2015,Willis2013}.
However, its extension to dynamical systems with huge degrees
 of freedom such as turbulent flows  has just come to be considered
recently and is still  one of challenging issues\cite{Kreilos2013}.

As a tractable and simple model representing localized
turbulence, we deal with a two-dimensional flow in doubly-periodic box
forced by a single monochromatic external force called Kolmogorov flow.
Kolmogorov flow  has been widely examined for a longtime to understand
mainly mathematical aspects of Navier-Stokes flow such as  cascades
 of supercritical bifurcations to turbulence
\cite{Gallet2013,Chandler2013,Lucas2014c,Lucas2015a,Sivashinsky1985,Marchioro1987,Meshalkin1961}.
Recently, spatially-localized dynamical states and its dynamical
properties have been reported\cite{Lucas2014c,Lucas2015a}.
Solitary spatially-localized turbulent
states can exist and  even be isolated by introducing the flow
rate as a  control parameter in the direction
in which the Galilean invariance is broken by the forcing\cite{Hiruta2015}.

\begin{figure}
  \centering
  \includegraphics[width=\linewidth]{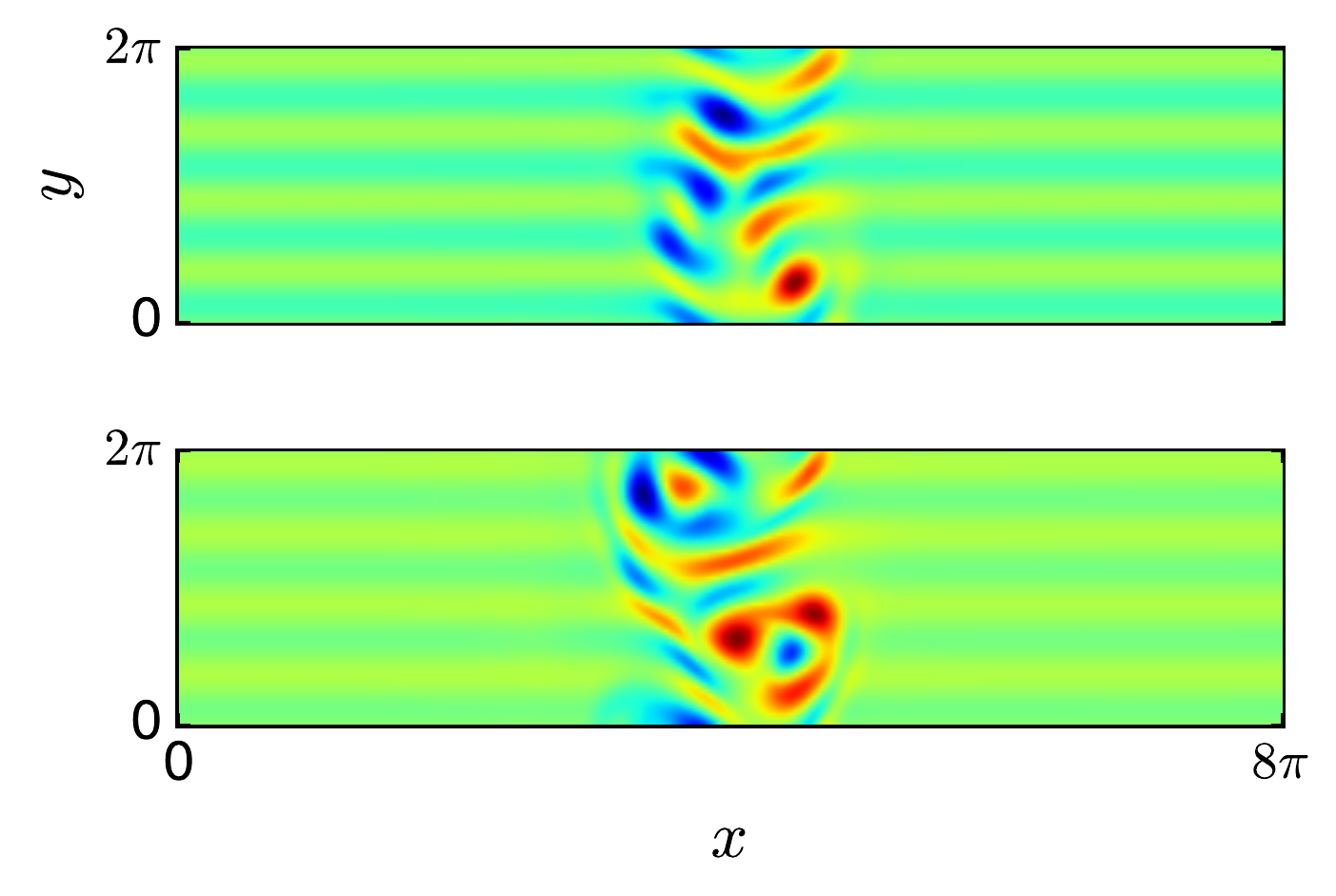}
  \caption{\label{fig:snap}
%  The snapshots of vorticity field $\omega$ of a spatially-localized
 %  traveling turbulence for $n=4$ and $\alpha=0.25$.
 Snapshots of vorticity field $\omega(\bm{x},t)$ of a moving SLT
 for $n=4$ and $\alpha=0.25$: for Re=26.75 and $U_y$=0.933 (top panel)
 the SLT begins to switch its moving direction; 
 for Re=50 and  $U_y$=1.46 (bottom panel).}
 \end{figure}

In this paper, we investigate novel translational motion of an SLT.
In two-dimensional Kolmogorov flow at moderate values of Reynolds
number and the flow rate,
an SLT as shown in FIG.\ref{fig:snap} moves with a nearly constant
speed sustaining its  direction for a long time and
suddenly and intermittently turns around  as shown in FIG.\ref{fig:slice}.
Our motivation is to clarify the relationship  between this translational
motion and the internal dynamics of a single SLT.

The rest of paper is organized as follows:
Sec.\ref{sec:setting} is  devoted to define and characterize
 the flow system. We introduce a co-moving frame to decompose
 an SLT into its spatial translation and internal dynamics.
 In sec.\ref{sec:exp}, the coarse-grained motion of the center of
 an SLT is examined. In sec.\ref{sec:int}, we try to describe the motion
 of the center with representative variables of the internal dynamics
 of the SLT in the co-moving frame. Concluding remarks are presented
in the final section.

\section{Governing Equation and setting\label{sec:setting}}
We focus on  two-dimensional (2D) Kolmogorov flow which is 2D flow
sustained by a steady sinusoidal force.
The velocity field $\bm{u}=(u_x, u_y)$, where the subscripts $x$ and $y$
denote the directions  parallel and perpendicular to the force, is
governed by the following 2D Navier-Stokes equation in doubly periodic domain $(x, y) \in [0,2\pi/\alpha]\times[0,2\pi]$:
\begin{align}
  \partial_t \bm{u}+(\bm{u}\cdot \bm{\nabla})\bm{u}  &= -\nabla p +\frac{1}{\rm Re} \nabla ^2 \bm{u}+\sin(ny)\bm{\hat{x}}, \label{eq:ns1}\\
  \bm{\nabla} \cdot\bm{u}&=0. \label{eq:ns2}
\end{align}
Here, the pressure $p$ is doubly periodic and  $\alpha$, Re, $n$ and
$\hat{\bm{x}}$ denote the aspect ratio of the rectangular domain,
 Reynolds number, the wave number of the external sinusoidal force and
 the unit vector in the $x$-direction, respectively.\\
The average flow rate in $y$-direction denoted by  $U_y$ is a
conserved quantity and  controls  the nature of the flow \cite{Hiruta2015}:
\begin{align}
  U_y=\frac{\alpha}{4\pi ^2}\int_{0}^{2\pi/\alpha}dx\int_{0}^{2\pi}dy u_y=\langle u_y \rangle _{xy}.
\end{align}

Direct numerical simulation (DNS) solves the following equation for the
vorticity,  $\omega=\partial_x u_y -\partial_y u_x$ with the
pseudo-spectral method for spatial
discretization using the tow-thirds rule for dealiasing and the 2nd order Runge-Kutta (Heun) method for time evolution.
\begin{align}
  \partial_t \omega+(\bm{u}\cdot \bm{\nabla})\omega  &= \frac{1}{\rm Re} \nabla ^2 \omega-n\cos(ny).\label{eq:vor}
\end{align}
The time and spatial resolutions used for DNSs are $2\times 10^{-3}$ and
128 points per $2\pi$, respectively.

This system has the following fundamental symmetries for $U_y\neq0$:
\begin{align}
  \mathcal{T}_l:\omega(x, y)& \rightarrow \omega(x+l, y)\, \, \left( 0\leq l < \frac{2\pi}{\alpha} \right), \\
  \mathcal{S}:\omega(x, y)& \rightarrow -\omega(-x, y+\frac{\pi}{n}).\label{eq:simR}
\end{align}
Here, $\mathcal{T}_l$ is a continuous translational symmetry in $x$-direction,
and $\mathcal{S}$ is a discrete shift-and-reflect symmetry
which is represented by cyclic group of order $2n$.
We also use these two symbols to denote actions on states of
a flow as long as there is no misunderstanding.

There are two main control parameters in 2D Kolmogorov flow: Re
and the flow rate  $U_y$. Note that for most researches on 2D Kolmogorov flow,
 $U_y$ is fixed to 0.
 Since we are interested in the relationship between the motion and  the
 internal turbulent dynamics of a single SLT,
 we limit Re to two values relatively close to the critical Re at which
 a moving SLT begins to switch its moving direction in $x$:
 ${\rm Re}=26.75$ and $50$ which are slightly and relatively higher
 than the critical Re.
 For a single SLT to exist in the box, the mean flow rate $U_y$ is set
to $0.933$ for ${\rm Re}=26.75$ and $1.46$ for ${\rm Re}=50$, respectively.
For the latter parameter set, the moving direction of an SLT contains
quick fluctuations as well as relatively slow and intermittent
 switching.
The other system parameters, $n$ and $\alpha$ are fixed to $(n,
\alpha)=(4, 0.25)$ in this paper.\\

In the lower Re case, the initial condition assigned is
 an unstable relative periodic solution (URO)
 which is a continuation solution of the stable
solitary relative periodic solution obtained in Ref.\cite{Hiruta2015}.
By the symmetry $\mathcal{S}$, this URO can have both positive and
negative velocities, $c_{\rm URO}=\pm0.02$, in $x$.
The period of the URO is  $\sim 60$ and characterizes the time scale of
the internal turbulent fluctuation.

Because of numerical errors in the initial condition,
this solution falls into an SLT in a few periods
and gets to switch  intermittently its moving direction. 
 Furthermore, around at $t\sim10^5$ 
 it suddenly ceases to move with a constant speed even on average and
 begins to hang around changing its moving direction quickly.
 This suggests that there exist several different types of SLT states: 
 A kind of transition from  traveling to standing SLT.
 However, we focus on the first (traveling) SLT state observed before
 the second transition.
 For Re = 50 and $U_y$=1.46, an SLT continues to travel with switching
 the direction for a long sustaining time at least $t\sim \mathcal{O}(10^6)$
 which is not affected by initial conditions with a single SLT.

 We introduce a frame system to separate the motion %and
 from the internal turbulent dynamics of each SLT.
 Here the motion of an SLT stands for the evolution in a coarse-grained
 time  of a point representing the location of the SLT. We call this
 point the center of the SLT.
 An SLT travels both in $x$ and $y$ directions:
 the translation in $y$ is mainly caused by the mean flow rate and
 the vortex pair constituting the SLT but  the translation in $x$
 is derived from the internal turbulent dynamics.
 Therefore we will give our attention to the motion in $x$ of an SLT.

 This frame system is an extension of Galilean transformation and 
 is defined by formally  applying a time-dependent translational symmetry:
 \begin{align}
  \hat{\omega}(x, y, t)=\mathcal{T}_{l(t)}\omega(x, y, t)=\omega(x+l(t),
  y, t), \label{def:homega}
 \end{align}
 where $l(t)$ is a time dependent shift in $x$-direction.
 We call the case of $l(t)=0$ the laboratory frame
 and the case of $l(t)=-X(t)$ the co-moving frame
 where $X(t)$ is an approximate or coarse-grained location of the
 SLT but its definition includes some ambiguity originated from
 the internal turbulent fluctuation.
 We define $X(t)$ by setting the phase of the first Fourier mode of
 the vorticity field with $k_x=\alpha$ and $k_y=0$ to $\pi/2$.
 Hereafter, we call $X(t)$ the center of the SLT.
 Note that if the velocity of the center of the SLT, i.e.,
 $V(t)=dX(t)/dt$ is not a constant, the dynamics of the SLT
 even in the co-moving frame is coupled with the motion of the SLT.

\begin{figure}
  \centering
  \includegraphics[width=\linewidth]{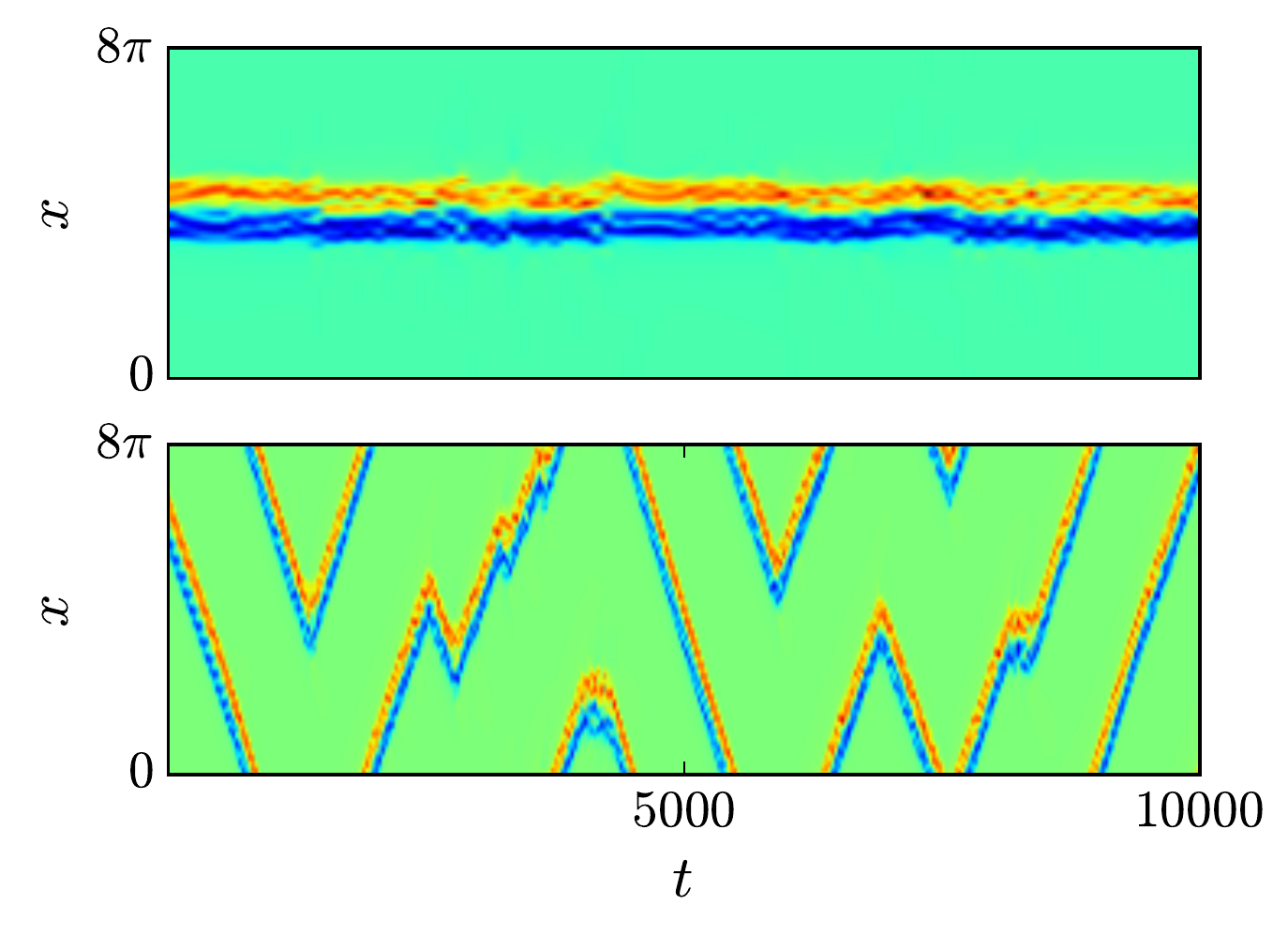}
  \caption{Time evolution of vorticity averaged in $y$-direction
 $\Omega(x,t)$ for Re=26.75 and $U_y=0.933$
 in the laboratory frame (top) and the co-moving frame (bottom).
  \label{fig:slice}
 }
\end{figure}

\begin{figure}
  \centering
  \includegraphics[width=\linewidth]{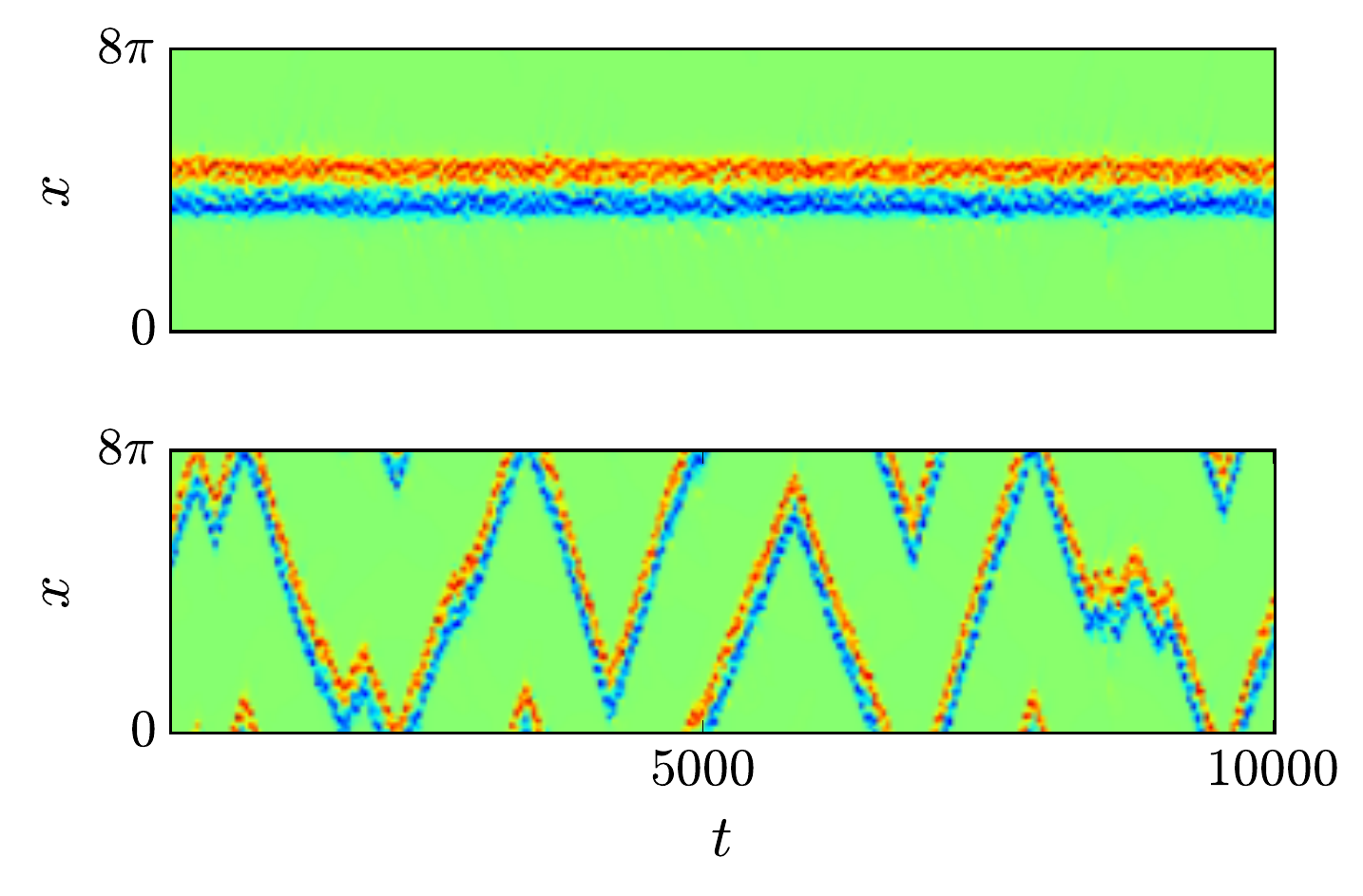}
  \caption{
  The same as FIG.\ref{fig:slice} except for Re=50 and $U_y=14.6$.
  \label{fig:slice2}
 }
\end{figure}

 Moreover, we expect (not guaranteed) that
 the center of the following average vorticity $\Omega(x,t)$ 
 stays around the same positions  in the co-moving frame:
 \begin{align}
  \Omega(x,t)=\frac{1}{2\pi}\int_{0}^{2\pi}dy \omega(\bm{x}, t).
  \label{eq:avevor}
 \end{align}
 This method has been adopted for 
 one-dimensional PDE and three-dimensional turbulent pipe
 \cite{Budanur2015,Willis2016}.
 Both in the laboratory and co-moving frames, 
 the time evolution of $\Omega(x,t)$ is shown
 in FIG.\ref{fig:slice} and FIG.\ref{fig:slice2}.

In the laboratory frame, sudden and intermittent changes of
the moving direction of the SLT can be observed. 
In the co-moving frame, the SLT stands around the same position 
in time with small fluctuation.
 This enable us to make a decomposition into the motion of
the SLT, i.e., $X(t)$  and  the internal turbulent dynamics,
$\hat{\omega}(\bm{x}, t)$ defined in (\ref{def:homega}).

\section{motion of moving turbulence\label{sec:exp}}
We focus on the nature of the switching of the moving direction.
Because even the coarse-grained center of an SLT, $X(t)$,  still
fluctuates in a time scale of the order of the internal dynamics of the SLT, 
the intervals between adjacent reverses of the moving direction denoted by
$\Delta t$, i.e., the residence time  are evaluated with
 a velocity 
 averaged over an interval $T$  defined as
\begin{align}
  \overline{c}_T(t)=\frac{1}{T}\int_{t}^{t+T}dt' \frac{dX}{dt'}(t')=\frac{X(T+t)-X(t)}{T}.
\end{align}
The velocity $\overline{c}_{T=+0}$ crosses  zero even if an SLT moves in
the same direction in the coarse-grained scale
because the average vorticity $\Omega$ stays around the same position
but strongly fluctuates in the translational direction especially
in the higher Re case as shown in FIG.\ref{fig:slice} and FIG.\ref{fig:slice2}.
To detect the direction reversal in a coarse-grained time,
 we set $T=100$, which is longer than a typical time scale of the internal
 turbulence dynamics.
 This typical time scale is $\sim 60$ and of the order of the period of
 the URO adopted as the initial condition.
 The subscript $T$ is omitted hereafter for simplicity.

\begin{figure}
  \centering
  \includegraphics[width=\linewidth]{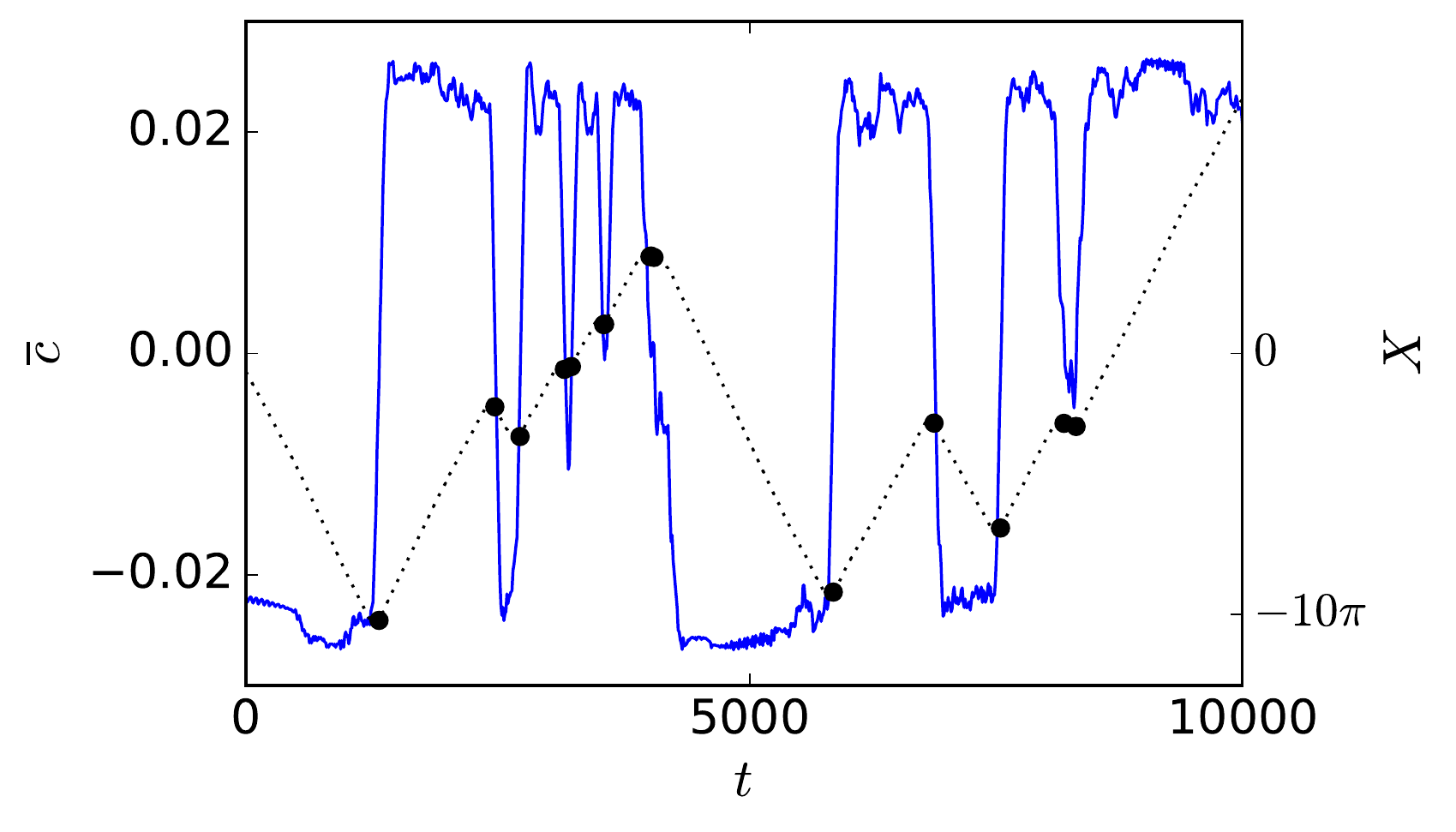}
  \includegraphics[width=\linewidth]{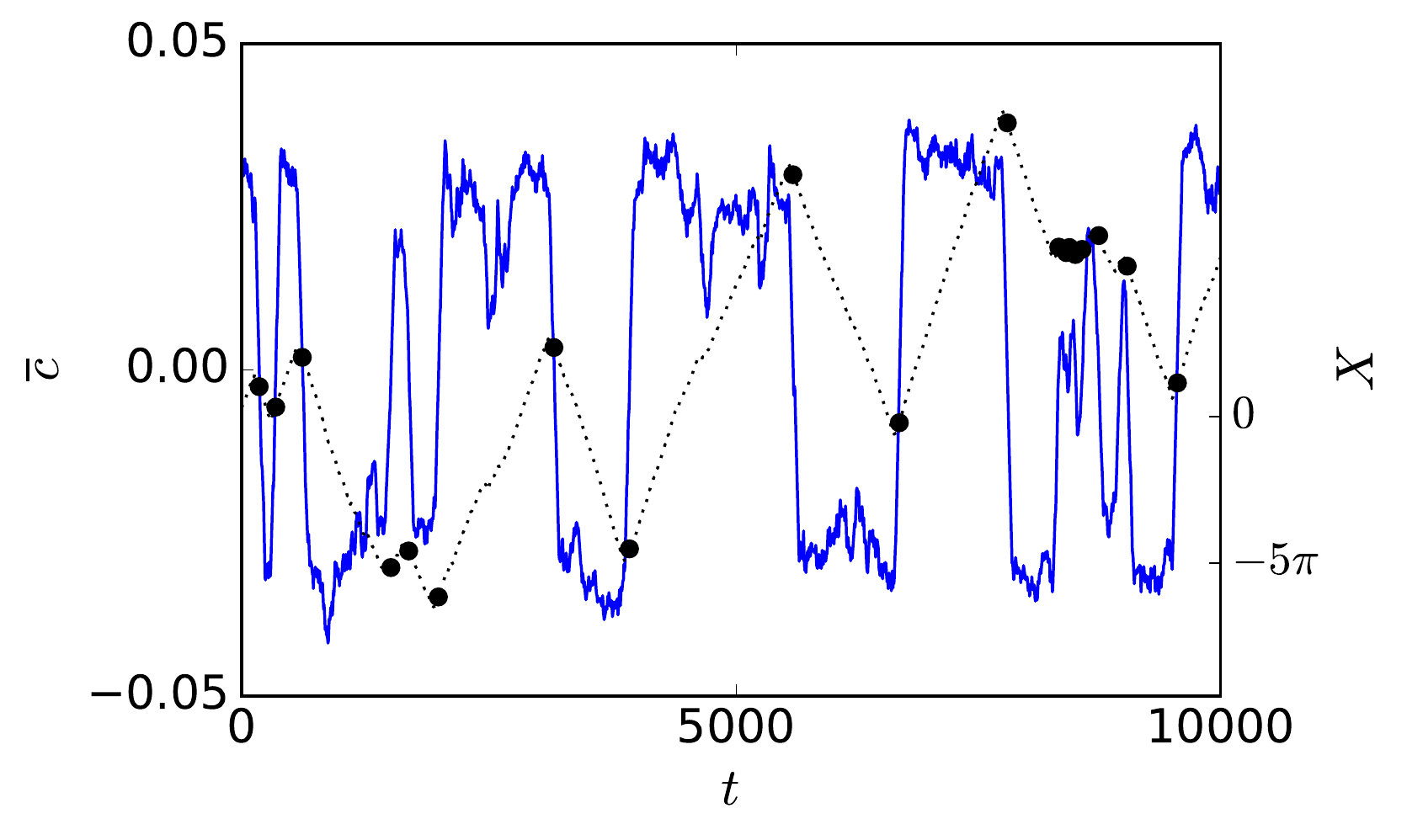}
  \caption{\label{fig:speed}
  Time evolutions of the average velocity $\overline{c}$ (blue solid
 line) and the central position of SLT $X$ (black dotted line).
  Large black dots denote direction reversals at 
   Re=26.5 (top panel) and 50 (bottom panel).
 }
\end{figure}

 The evolutions of the center $X(t)$ and the average velocity $\overline{c}(t)$
 are shown in FIG.\ref{fig:speed}.
The average velocity $\overline{c}(t)$ takes roughly two values,
i.e. $\pm|c_{\rm max}|$  and a direction reversal occurs  when
$\overline{c}(t)$ crosses zero.
In this sense, the average velocity $\overline{c}(t)$ is an adequate 
variable to detect direction reversals.
This also  suggests that at least there are  two (that is, twin)
 unstable invariant sets with $\pm|c_{\rm max}|$ about one of which
the SLT wanders and the direction reversal corresponds
to switching between the stays around these sets.
We expect that these invariant sets are close to the twin URO one of
which is adopted as the initial condition.

\begin{figure}
  \centering
  \includegraphics[width=0.49\linewidth]{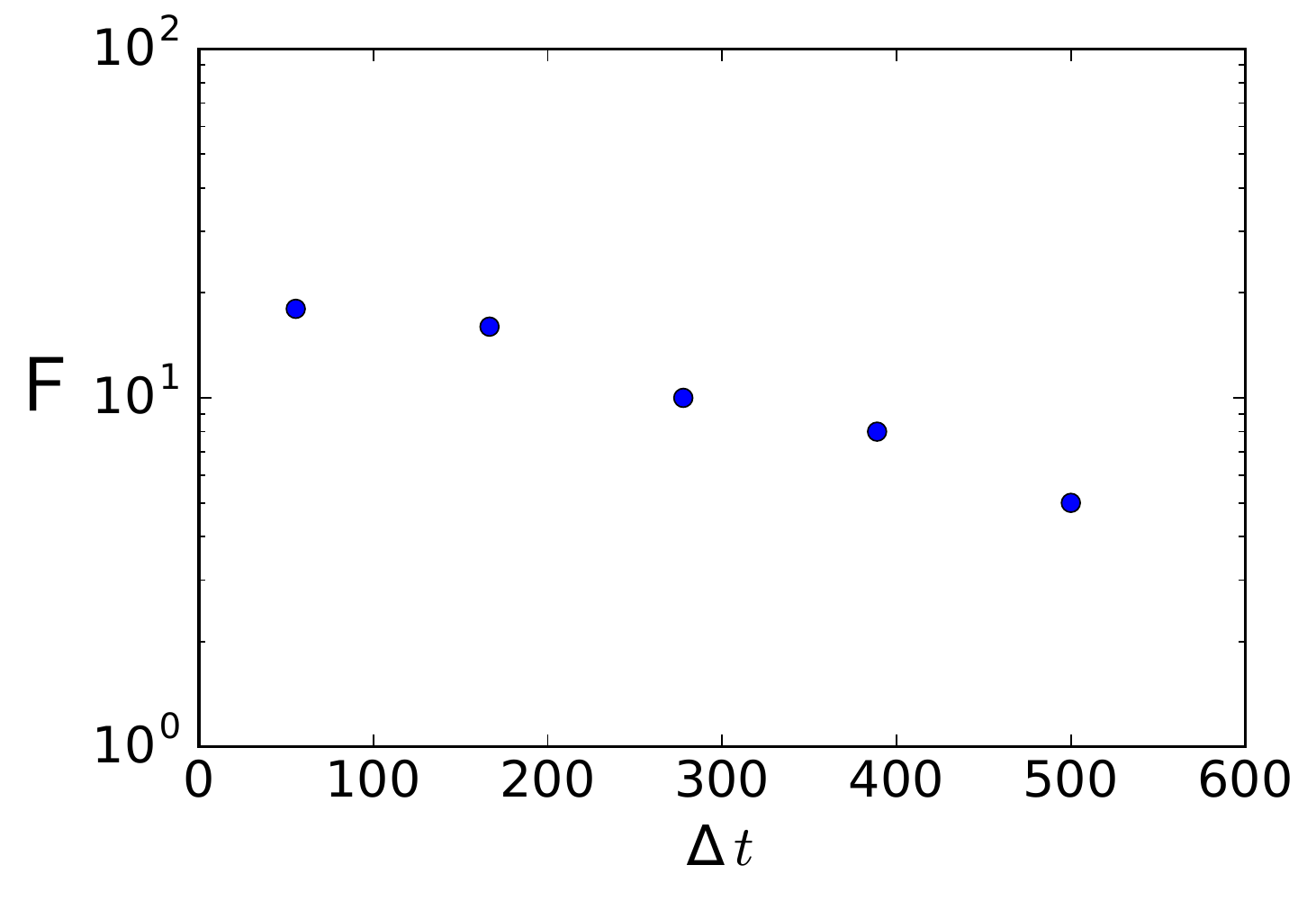}
  \includegraphics[width=0.49\linewidth]{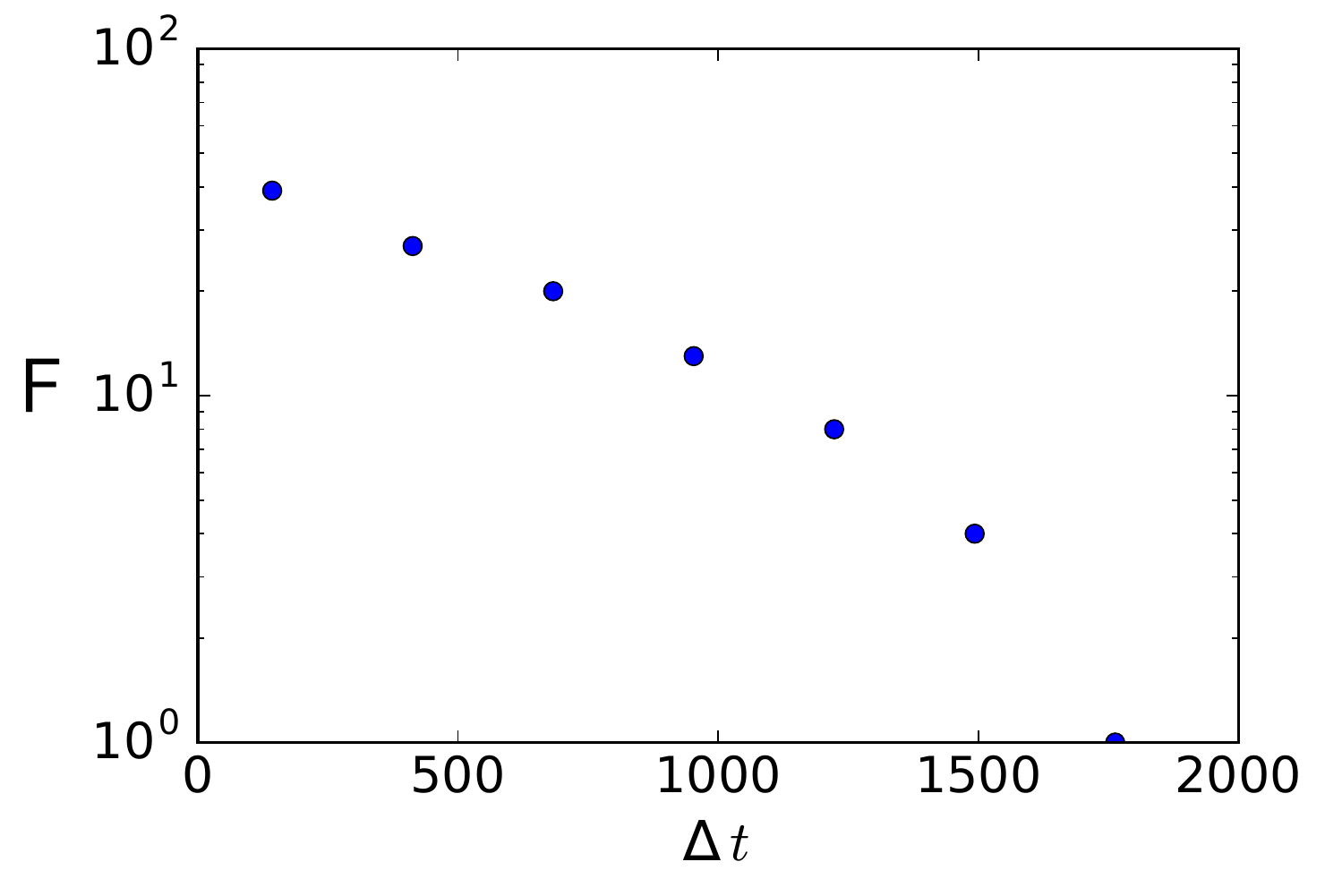}
  \caption{$F(\Delta t)$ for 
  (Re, $U_y$)=(26.75,0.933) (left panel) and
  (Re, $U_y$)=(50,1.46) (right panel).
  \label{fig:cdf}
  }
\end{figure}

The histogram of the  number of the residence time $\Delta t$ larger than $t$
denoted by $F(\Delta t>t)$ is shown in FIG.\ref{fig:cdf} and exponential decays
 are suggested for both Reynolds numbers.
 Note that the residence time $\Delta t$ is sufficiently longer than
 the average time $T=100$, although  in the lower Re case  enough amount of samples
can not be taken because of its finite lifetime as mentioned
in Sec.\ref{sec:setting}.

The exponential-decay tendency observed in $F(\Delta t >t)$ 
reminds us of a random telegraph signal which is produced by the Poisson
process\cite{Anishchenko2003,Varon2016}. This suggests that the aforementioned twin invariant sets
 corresponding to SLTs with the positive and negative velocities in $x$
have complicated structures different from simple spiral chaos such as
Lorenz attractor. It also seems to support the simplified picture
mentioned above of the phase space in which the motion of a single SLT
is embedded. Therefore we should carefully select a well-acted
projection to describe the trajectory in a phase space.

\section{Relation to internal turbulent dynamics\label{sec:int}}
In this section, we try to describe the motion of a single SLT
in the phase space in relation to internal turbulent dynamics.
To begin with,
we introduce some coarse-grained variables that characterize the
asymmetric nature in $x$  of the internal turbulent dynamics
observed in the co-moving frame based on $\mathcal{S}$
which allows an SLT to travel both to the negative and
positive directions in $x$.
 The first one is the following quantity evaluated simply by the maximum and
 minimum values of vorticity:
 \begin{equation}
  s(t)=\max \omega(\bm{x},t) -|\min\omega(\bm{x},t)|.
 \end{equation}
 Since the transformation $\mathcal{S}$, $x\to -x$ and  $y \to y+\pi/n$,
 changes the sign of the vorticity $\omega(\bm{x},t)$, the sign of $s(t)$
 also changes as follows:
\begin{equation}
\mathcal{S}s(t) = -(\min \omega(\bm{x},t)) -|-\max\omega(\bm{x},t)|=-s(t). 
\end{equation}
 The variable $s(t)$ evaluates the degree of the asymmetry of the
 vorticity distribution of an SLT. This asymmetry is the origin of
 the antisymmetry of $s(t)$ and thus closely related  to
 the direction reversal.

\begin{figure}
  \centering
  \includegraphics[width=\linewidth]{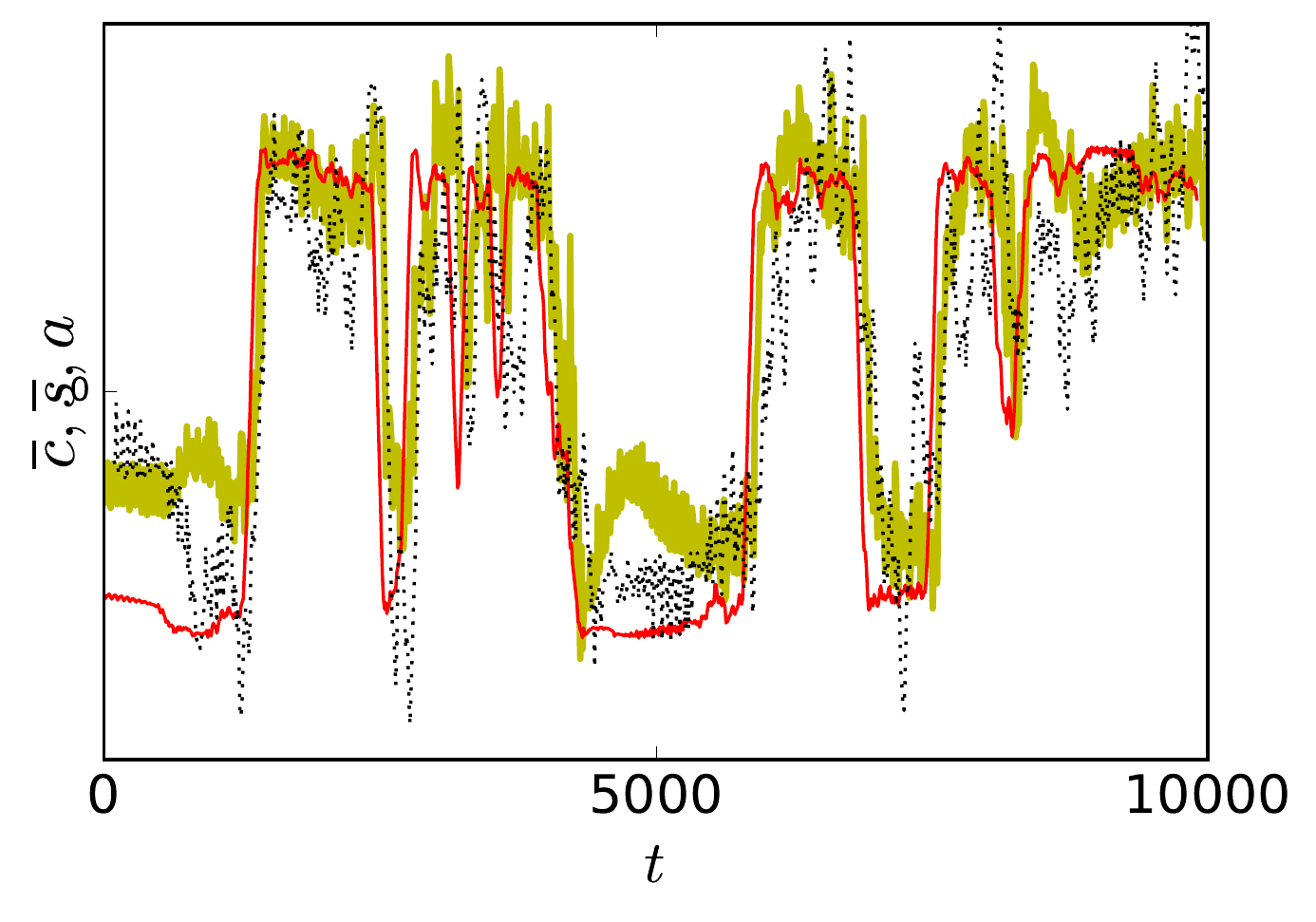}
  \includegraphics[width=\linewidth]{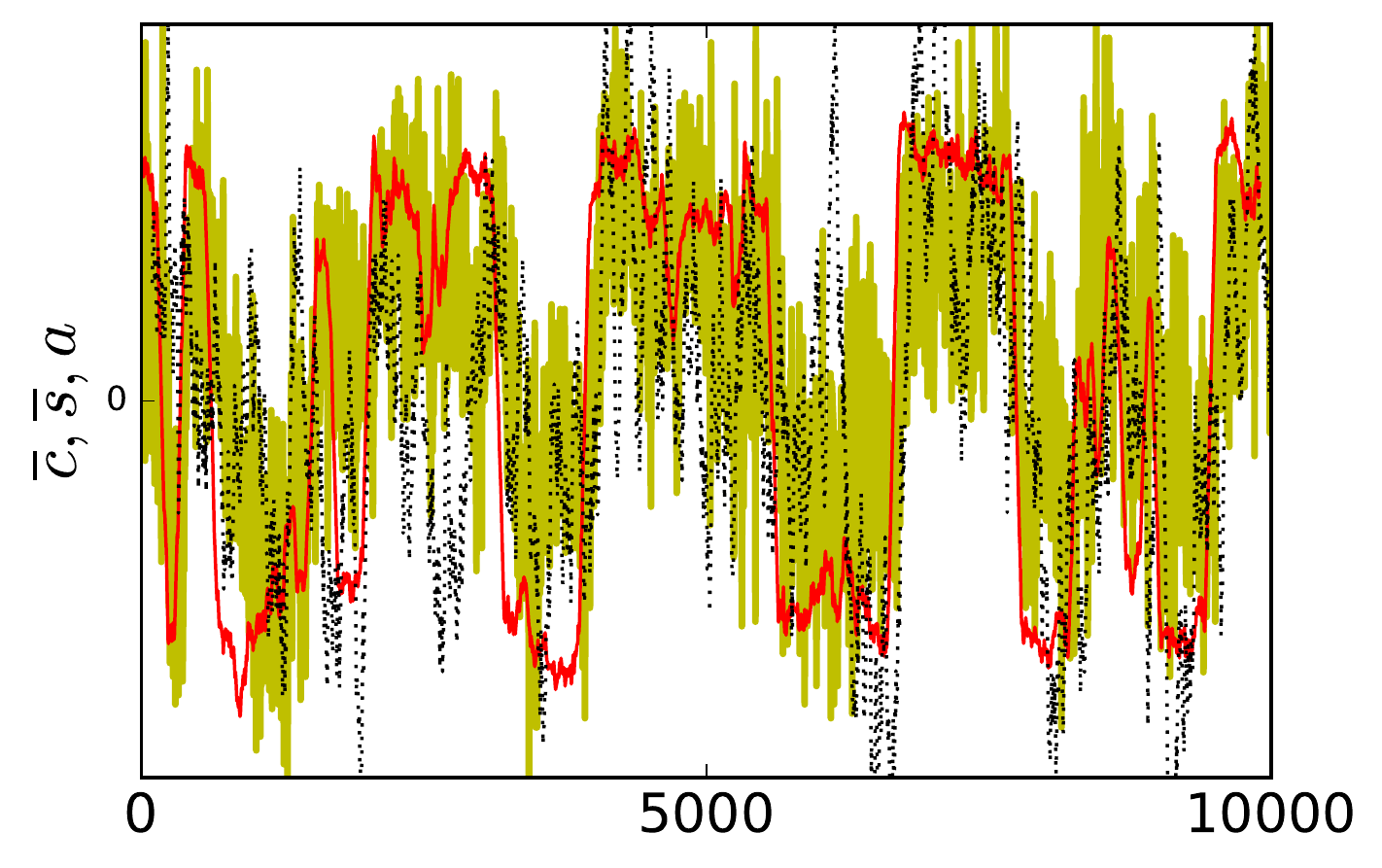}
  \caption{\label{fig:maxmin}
 Time evolutions of $\overline{c}$ (red thin solid line),
 $\overline{s}$ (black dotted line) and $a=a_p-a_n$ (yellow thick solid
 line): for Re=26.75 and $U_y=0.933$  (top panel) and
 for Re $=50$ and $U_y=1.4$ (bottom panel).
 }
\end{figure}

  %However,
 Since the maximum and minimum of $\omega(\bm{x},t)$
 fluctuate quickly in time like $\overline{c}_{T=+0}$,
 the average or coarse-grained  $\overline{s}$ should be also introduced:
\begin{align}
 \overline{s}_T=\frac{1}{T}\int_{t}^{t+T}dt's(t).
\end{align}
The subscript $T$ of $\overline{s}_T$ is  omitted hereafter for simplicity.
As shown in FIG.\ref{fig:maxmin},
$\overline{s}$  correlates adequately with the moving direction
of an SLT in the two Re cases. However, they fluctuate more strongly
than $c_T(t)$ and this tendency is enhanced in the higher Re case.
This suggests that  the internal turbulence controls the motion of a single SLT.
However, the correlation between $\overline{s}_T(t)$ and $\overline{c}_T(t)$ is
not sufficient enough  for  $\overline{s}_T$ to be used for quantitative
description of the motion of the SLT.

We next introduce another variable representing a kind of
the distance from these invariant sets more quantitatively.
The vorticity field in the co-moving frame
$\hat{\omega}(\bm{x},t)$ is projected onto the
two fields defined by the average under the condition that
the traveling direction is positive or negative, respectively.
The negative mean state $\phi_n$ and the positive mean state $\phi_p$
are defined numerically as follows:
\begin{align}
 \phi_n(\bm{x},t) &=\langle\hat{\omega}(\bm{x},t)\rangle_{\overline{c}(\hat{\omega})<0} \label{eq:mean_n}, \\
  \phi_p(\bm{x},t)&=\langle\hat{\omega}(\bm{x},t)\rangle_{\overline{c}(\hat{\omega})>0} \label{eq:mean_p}, \\
\end{align}
where the bracket $\langle\ \rangle$ and its subscript denote an
ensemble average in the co-moving frame and the condition under
which the average is calculated, respectively.
These two fields are expected to approximate the twin invariant sets.

The projections onto $\phi_n$ and $\phi_p$ are carried out
 with the internal product $\langle\phi|\omega\rangle$ between real
 functions as follows:
\begin{align}
 a_n(\hat{\omega})&=\frac{\langle\phi_n|\hat{\omega}\rangle}{||\hat{\omega}||^2},\\
  a_p(\hat{\omega})&=\frac{\langle\phi_p|\hat{\omega}\rangle}{||\hat{\omega}||^2},\\
\langle f|g\rangle&=\frac{\alpha}{4\pi^2}\int dxdy f(x,y)g(x,y),\\
  ||f||&=\sqrt{\langle f|f\rangle}.
  \end{align}
  The difference between the coefficients denoted by $a(t)=a_p(t)-a_n(t)$
  also evaluates the asymmetry of an SLT and is expected to indicate
  the direction of the motion,
  because the moving direction of the URO is determined by
  the asymmetry of the vorticity field and an SLT seems
  to stay around  one of the invariant sets close
  to the corresponding URO.
  As shown in FIG.\ref{fig:maxmin}, $a(t)$ reproduces roughly
  the switching process better than $\overline{s}(t)$.
  However, in the higher Re case, both the two quantities,
 $\overline{s}(t)$ and $a(t)$, which are  coarse-grained representatives
  of the internal turbulent dynamics, tend to be less able to follow
  the average velocity $\overline{c}(t)$,
  although the two moving states with the velocities $\sim \pm|c_{\rm max}|$ 
  and the direction reversal are still clearly identified.
   This suggests that even though the twin or multiple invariant sets
   are still discriminated  clearly by the moving direction,
   the number of variables or the dimension of the  phase space
   required to describe the internal turbulent dynamics
   which is directly related to the coarse-grained motion of an SLT
   increases with Re.

\begin{figure}
  \centering
  \includegraphics[width=\linewidth]{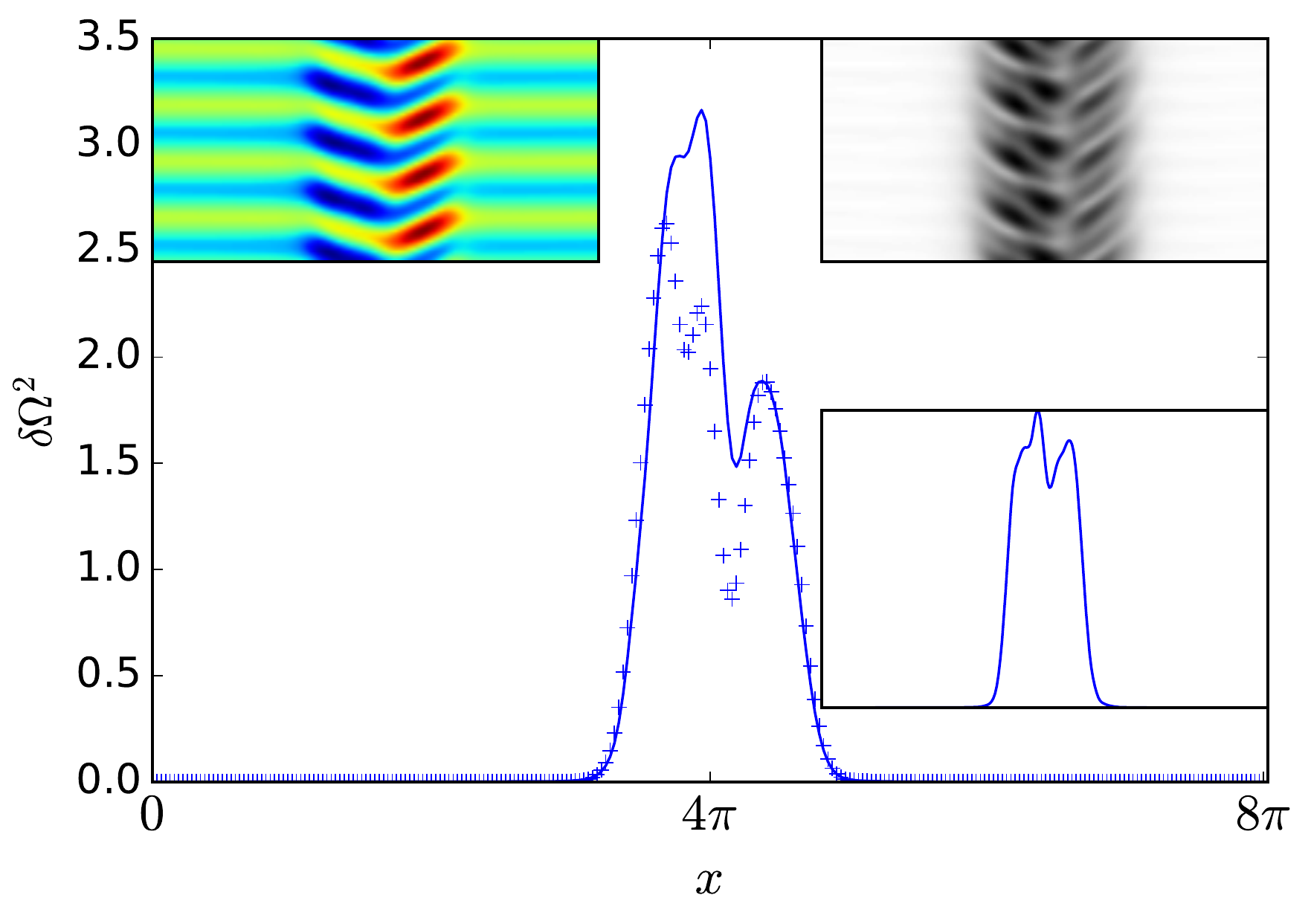} 
  \caption{\label{fig:std} %Fluctuation of vorticity fields $\delta\Omega_n^2(x)$ for Re=26.75 and $U_y$=0.933
 Dispersion of vorticity field  $\delta\Omega_n^2(x)$
  defined in Eq.(\ref{eq:fluct}) for Re=26.75 and $U_y$=0.933.
 %equation (\ref{eq:fluct}).
% Ensemble take when $\overline{c} < 0$: SLT is moving to left.
 The average is taken under the condition $\overline{c} < 0$: 
 % take when $\overline{c} < 0$: 
  Solid line for DNS and plus signs for URO.
  Left inset shows  a mean vorticity field $<\omega>_{\overline{c}<0}$
  and right upper inset shows  dispersion field $\delta\omega^2_n(x,y)$.
  Each of upper insets shows only a main part of the field.
  Right lower inset shows $\delta\Omega_n^2(x)$ for Re=50 and $U_y$=1.46.
% Solid line indicates from DNS and dotted line indicates URO.
%  Left inset indicate mean vorticity fields $<\omega>_{\overline{c}<0}$ and right inset indicate fluctuation field $\delta\omega^2$.
%  Each of insets show only a main part.
 }
\end{figure}

This difficulty is inherited by their higher order moments.
To see this, we define the dispersion, the second order fluctuation
of the vorticity field under the condition $\overline{c}<0$ as follows:
\begin{align}
 \delta\Omega_n^2(x) &=\int_0^{2\pi} dy \delta\omega^2_n(x, y), \label{eq:fluct} \\
  \delta\omega_n^2(x, y)&=\langle\omega^2(x, y)\rangle_{\overline{c}<0}-\langle\omega(x, y)\rangle_{\overline{c}<0}^2.
% \delta\omega_n^2(x, y)&=<\omega^2(x, y)>_{\overline{c}<0}-<\omega(x, y)>_{\overline{c}<0}^2.
\end{align}
 Figure \ref{fig:std} shows the dispersion $\delta\Omega_n^2(x)$ in the
 both Re cases  and is compared with that of the URO.
 Since the SLT travels to the left, i.e., $\overline{c}<0$,
 the fluctuation on the left side or the front  of the SLT is
 stronger than that of the right side or the back front, while the
 absolute value of the average vorticity is larger on the back front 
 than on the front.
 As shown in FIG.\ref{fig:std},
 this characteristics of the vorticity fluctuation  is  shared
 with a left traveling URO of $c_{\rm URO}=-0.02$ which is
 a periodic solution in the co-moving frame though the asymmetry
 of the dispersion of the URO is  weaker than that of the SLT. 
 This also  supports  the simplified picture of the phase space
 constituted by several twin unstable invariant sets each
 of which might correspond to a one-way traveling SLT.
 The asymmetry of the vorticity fluctuation is observed in 
 other traveling localized states or invasion-fronts of turbulence.\cite{Barkley2015,Teramura2016}
  In the higher Re case, the asymmetry of vorticity fluctuation
 is weaker than that in the lower Re case.
 Therefore we need variables more susceptible to geometrical or
 temporal characteristics of the SLT to resolve
 the internal turbulent dynamics.

\section{Concluding remarks\label{sec:con}}
  We have found that a single spatially-localized turbulence (SLT)
  exists stably and travels in $x$ switching the moving direction
  randomly and intermittently.
  By introducing the coarse-grained center $X(t)$ and traveling velocity
 $\overline{c}_T(t)$ of the SLT,
 we have characterized the traveling motion and these switching events
 in the coarse-grained time scale.
 Even in the relatively high Re case, 
 $\overline{c}_T(t)$ takes roughly two values $\pm|c_{\rm max}|$ and the residence
 time $\Delta t$ is thought to obey the exponential distribution,
 which suggests that  $\overline{c}_T(t)$ can be approximated
 by a random telegraph signal. At least (several) twin attracting
 invariant sets, each of which  corresponds to a one-way traveling SLT
 and may be close to  unstable relative periodic solutions (URO),
  are embedded in the attractor of the moving turbulence. 
  Since we expect that like a self-propelled particle the motion of a single
  SLT is controlled by the characteristics of the internal turbulence,
 the time evolution of the flow is  decomposed into the coarse-grained
 motion of the center of the SLT and  the accompanying turbulent field
 by defining the co-moving frame with $X(t)$.
 We have introduced  two  coarse-grained variables $\overline{s}_T(t)$
and $a(t)$ characterizing the asymmetry in $x$ of the internal
turbulence:  $\overline{s}_T(t)$ simply estimates the asymmetry of the vorticity distribution
 of the SLT and $a(t)$ represents the difference between the approximated
 distances from the twin unstable sets one of which is projected onto
 the other by the discrete shift-and-reflect symmetry $\mathcal{S}$.
 In the lower Re case both the variables follow $\overline{c}_T(t)$ sufficiently.
 However, in the higher Re case, though the switching events are detected
 well and  the latter seems to work relatively better than the
 former, they fluctuate more significantly than $\overline{c}_T(t)$.
 This implies that with Re the structure in the phase space of each 
 one-way SLT gets complicated  and thus more variables or
 dimensions are required to resolve it. 
 The difficulty in the higher Re case might partly come from an arbitrary
way of decomposition into position and internal dynamics.
In other words, the fluctuations of quantities in co-moving frame are
affected by the definition of the center of position, which might be
solved by introducing a  proper co-moving frame.
However, such a frame could not be  found a prior in general.

Since the switching processes are detected sharply and can be well
approximated by a random telegraph signal even in the higher Re case, 
the twins must be  still separated clearly in the attractor of a single
SLT. 
In this paper, we have not dealt with the mechanism of the reversal of the
moving direction and its relationship with the internal turbulence but
 focused on the characterization of the coarse-grained motion of a
SLT and its internal turbulent.
It is easy to make a Langevin model which can reproduce the
stochastic nature of the switching events.  However, we are now
rather trying to study a deterministic model of the switching
process in relation to the internal turbulence focusing on 
the structure of the attractor.

Invariant solutions should play a key role in more reliable description
of states at higher Re.
It is expected that a fixed point like URO embedded in each of the twin,
 i.e., a pair of chaotic attracting sets  mimics the average quantity for each directions in FIG.\ref{fig:std}.
The exponential-like decay of the residence time $\Delta t$ 
suggests that the switching  between the twin occurs randomly like a
 random telegraph signal and thus the way to  connect the twin is
 complicated but expected to tightly relate to the internal turbulent dynamics.
These invariant solutions also will help us to attain  more quantitative and
precise  understanding of SLTs.

This type of intermittent switching can be observed in other flows: 
reversal of Large Scale Circulation in a steady forced flow
\cite{Sommeria1986,Mishra2015}
and thermal driven flow\cite{Sugiyama2010,Ni2015}.
The approach  based on a low dimensional model derived by Galerkin method
is useful to study such a transition\cite{Shukla2016}.
We expect that by this approach with the invariant solutions
mathematical models representing a moving SLT can be developed.

 Since as mentioned above the residence time is suggested to obey an
 exponential distribution, the switching process seems to be
 Poisson process like interval statistics.
 However, the total number of events obtained are too little
 to decide a class of direction reversals.
 Therefore we should perform longer DNS repeatedly to
 obtain precise statistical aspect of direction reversals.

 It is interesting and important to study states consisting of
 a number of SLTs\cite{Hiruta2015}.
 Such a multiple SLT state in Kolmogorov flow also can contribute
 as one of  the most simple and tractable examples
 in the elucidation of turbulence transitions  observed in wall-bounded flows.
 However, our approach introduced in this paper needs some improvements
 in the definitions of  the coarse-grained quantities such as positions,
 velocities and  ones representing internal turbulent dynamics.

 Concerning subcritical turbulence transition as non-equilibrium phase
 transition, "moving" SLTs that are observed there may affect the
 determination of critical exponents  and/or even a class of the
 transition itself.
 In fact, to do so  spatial and temporal intervals of
 laminar regions have been utilized, but fast moving SLTs
 might modify the distribution of such intervals which
 blurs the estimation of critical exponents.
 At least, the correlation length between SLTs can be much longer than 
 the length scale of the support of an SLT.
 In statistical physics, Mermin-Wagner theorem shows that no long-range
 order exists for systems at thermal equilibrium in two spatial
 dimension, e.g. XY model\cite{Mermin1966}. 
 However,  XY model consisting of moving elements such as Viscek model
 can have a non-zero order parameter and discontinuous phase transition
 occurs even in two spatial dimensions though Viscek model is
 a non-equilibrium system\cite{Vicsek1995,Toner1995,vicsek2012a}.
 This might suggest that subcritical turbulence transition does not
 necessarily belong to the universal class of directed percolation if 
 fast moving elements exist there.
\begin{acknowledgments}
 The authors thank Dr. Teramura for useful discussions and comments on our work.
\end{acknowledgments}
\bibliography{main}

\end{document}